\documentstyle[11pt,fleqn]{article}
\date{}
\def\picture #1 by #2 (#3){
  \vbox to #2{
    \hrule width #1 height 0pt depth 0pt
   \vfill
    \special{picture #3} 
    }
  }
\def\scaledpicture #1 by #2 (#3 scaled #4){{
  \dimen0=#1 \dimen1=#2
  \divide\dimen0 by 1000 \multiply\dimen0 by #4
  \divide\dimen1 by 1000 \multiply\dimen1 by #4
  \picture \dimen0 by \dimen1 (#3 scaled #4)}
  }
\begin{document}
\thispagestyle{empty}
\title{\bf Large Scale Structure by Global Monopoles and Cold Dark Matter}
\author{}
\maketitle
\begin{center}
{\Large \bf Leandros Perivolaropoulos}\\
\vspace{.5cm}
Division of Theoretical Astrophysics\\
Harvard-Smithsonian Center for Astrophysics\\
60 Garden St.\\
Cambridge, Mass. 02138 \\
{\it and}\\
Department of Physics \\
Brown University \\
Providence, R.I. 02912, U.S.A.\\
\vspace{.5cm}
\end{center}

\begin{abstract}
A cosmological model in which the primordial perturbations are provided by
global
monopoles and in which the dark matter is cold has several interesting
features.
The model is normalized by choosing its single parameter within the bounds
obtained
from gravitational wave constraints and by demanding coherent velocity f1ows of
about 600km/sec on scales of
 $50 h^{-1} Mpc$. Using this normalization, the model predicts the existence of
dominant structures
with mass $2\times 10^{16} M_\odot$ on a scale $35 h^{-1}Mpc$ i.e. larger than
the horizon at
$t_{eq}$. The magnitude of the predicted mass function in the galactic mass
range is in good
agreement with the observed Schechter function.\end{abstract}

\section{\bf Introduction}
\par
Recent observations have revealed the existence of nonlinear structures
on very large scales (up to $50 h^{-1}$Mpc). This observational fact has been a
challenge
for theoretical models during the last few years. In fact the currently popular
models
based on adiabatic primordial perturbations produced during inf1ation
have difficulty to account for such large structures while maintaining
successes on smaller scales.
It is therefore important to investigate alternative theories.
\par
A new class of such theories
based on primordial perturbations produced by seeds rather than a superposition
of waves with random phases has attracted significant attention during the past
few years.
Physically motivated candidates for such seeds are topological defects
\cite{v85}.
Topological defects are localized energy concentrations which are predicted by
many
particle physics models to form during phase transitions in the early universe.
According to their geometry topological defects appear in the form of
monopoles (stable pointlike defects), cosmic strings (linear defects) domain
walls
(planar defects) and textures (unstable pointlike defects).
The vast majority of literature had until recently focused on {\it gauged}
cosmic
strings: linear defects interacting with short range forces. It was later
realized\cite{t89}
however that {\it global} defects interacting with long range forces, have two
important
advantages over  {\it gauged} defects:
\begin{enumerate}
\item  They generically have a larger correlation length during their evolution
and
can therefore seed structures on larger scales than gauged defects.
\item
 Their long range Goldstone field can provide efficient annihilation mechanisms
that prevent them from dominating the energy density of the universe and
lead to a scale invariant distribution: the scaling solution.
\end{enumerate}
Thus, for example,
monopole-antimonopole annihilation makes global monopoles consistent with
standard cosmology \cite{rb90}, in contrast to the case of gauged monopoles.
These advantages have led to recent extensive study of the unstable pointlike
defect
-global texture- with very encouraging results\cite{gpstg91}.
\par
In this letter we consider a model in which the large scale structure is seeded
by
stable pointlike defects (global monopoles) and in which the dark matter is
cold.
Global monopoles may form in particle physics models where the breaking of a
global symmetry results in a vacuum manifold with topology $S^2$.
Such a breaking is realized in a model with the simple Langrangian density:
\begin{equation}
{\cal L} = {{1 \over 2} {\partial_{\mu}\vec\Phi}\cdot{\partial^{\mu}\vec\Phi}}-
{{1\over 4} \lambda ({\vec\Phi}\cdot{\vec\Phi}-\eta^2)^2}
\end{equation}
where $\vec \Phi$ is a scalar field with O(3) global symmetry and the symmetry
breaking is
$O(3)\rightarrow O(2)$.
 Global monopoles are topologically nontrivial solutions
of the above model. They are described by the spherically symmetric ansatz:
\begin{equation}
 {\vec \Phi}= f(r) {{\vec r} \over  r}
\end{equation}

Even though a solution for $f(r)$ is not known analytically, it is easy to
show using the equations of motion that its asymptotic behaviour
 is  $f(r)\rightarrow 1$
as $r\rightarrow \infty$ and $f(r)\rightarrow r$ for $r\rightarrow 0$.
Thus the energy momentum tensor may be approximated at large distances by
 \begin{equation}
 T^t_t\simeq T^r_r\simeq {\eta^2 \over r^2},\hspace{5mm}  T^\theta_\theta =
T^\phi_\phi \simeq 0 \end{equation}

Solving the Einstein equations with this $T^\mu_\nu$ leads to the global
monopole
metric outside of the monopole core \cite{bv89}
\begin{equation}
 ds^2=dt^2-dr^2-(1-8\pi G\eta^2)r^2(d\theta^2 + \sin^2\theta d\phi^2)
\end{equation}

Thus there is a spherical deficit angle $\Delta=8\pi G\eta^2$. Equivalently
any planar surface containing the monopole, has the geometry of a cone with
deficit
angle $\Delta$. A more detailed analysis shows that in addition to the deficit
angle the monopole induces an effective small {\it negative} mass M in its
spacetime \cite{hl90}.
The magnitude of this mass however is proportional to the scale of symmetry
breaking
$\eta$ and is too small to be relevant for structure formation.
On the other hand, the effects of the deficit angle are independent from
the distance to the monopole core and may therefore be important even for very
small
$\Delta$.
By obtaining the geodesic equations in  the spacetime (4) it may be shown that
a
monopole moving with velocity $v_m$ induces a velocity perturbation $\Delta v$
towards the line of its trajectory, to its surrounding matter.
Ignoring terms of O($v_{matter} \over v_{monopole}$) and O($v_{matter} \over
c$), the magnitude of the dominant component  of
$\Delta v$ is $\Delta v=4\pi G \eta^2
v_m \gamma_m$ and its direction is perpendicular to the monopole trajectory .
These primordial velocity perturbations may then grow to form galaxies,
clusters
and large scale structure. In what follows we obtain some of the predictions of
the
above described model. We make the assumption that
the monopoles move with relativistic velocities on straight line trajectories.
This assumption is justified due to the long range attractive monopole-
antimonopole forces\cite{bv89},\cite{p91}  which are expected to induce
relativistic
velocities.
\par
The structure of this paper is as follows: In the next
section we use the Zeldovich
approximation to obtain the growth of perturbations induced by global
monopoles.
 Then we obtain the peculiar velocities predicted by the model and we use this
result to normalize the single free parameter $\eta$ of the model. In section 3
we use the results of the Zeldovich approximation to obtain the spectrum of
nonlinear masses predicted by the model.We also obtain the predicted
mass function and we compare the galactic mass range with the observed
Schechter function.  Throughout the paper $h$ is the Hubble
constant in units of $100km/(sec\cdot Mpc)$ and we consider a spatially f1at
$\Omega=1$ universe.

\section {\bf The Zeldovich Approximation}
\par
We will use the Zeldovich approximation to calculate the growth of the velocity
perturbations induced by relativistic global monopoles. The calculation is
similar to
the one in the case of cosmic string wakes \cite{svbst87}, \cite{pbs90} with
the important difference
that in this case the geometry of each wake is cylindrical rather than planar.
Consider a particle
with unperturbed comoving distance q from the trajectory of a global monopole
(assumed to be a
straight line). Due to the initial velocity perturbation induced by the
monopole a comoving position
displacement $\Psi(q,t)$ will develop. Thus the physical distance $r_{\perp}$
from the monopole
trajectory may be written \begin{equation}
r_{\perp}(q,t)=a(t) (q-\Psi(q,t))
\end{equation}
where $a(t)$ is the scale factor of the universe.
Also, in the Newtonian approximation we have
\begin{eqnarray}
{\ddot r_{\perp}} & = & -\bigtriangledown_{r_{\perp}}\Phi\\
{\bigtriangledown^2_{r_\perp}} \Phi & = & 4 \pi G \rho
\end{eqnarray}

Using (5), (6), (7) and linearizing in $\Psi$ leads to
\begin{equation}
{\ddot \Psi}+2 {{\dot a}\over a}{\dot\Psi}+3{{\ddot a}\over a}\Psi=0
\end{equation}

The velocity perturbation induced by a monopole at time $t_i$ is
$\Delta v=4\pi G v_m \gamma_m$. Therefore, the initial conditions to be used
for
the solution of (8) may be written in the comoving frame as:
\begin{eqnarray}
 \Psi(t_i) & = & 0,\hspace{1mm} {\dot\Psi}(t_i)=\Delta v({t_0\over
t_i})^{2\over
3},\hspace{5mm}t_i>t_{eq}\\
 \Psi(t_i) & = & 0,\hspace{1mm} {\dot\Psi}(t_i)=\Delta v({t_0\over
t_{eq}})^{2\over
3}({t_{eq}\over t_i})^{1\over 2},\hspace{5mm}t_i<t_{eq}
 \end{eqnarray}

The evolved comoving displcement $\Psi(t,t_i)$ with $t\geq t_{eq}$ is now
easily obtained
from (8), (9) and (10) and keeping only the growing mode it may be written as
 \begin{eqnarray}
\Psi_>(t,t_i) & = & {3\over 5}\Delta v \hspace{1mm}t_0^{2\over 3} t_i^{-{1\over
3}}t^{2\over
3},\hspace{5mm} t_i>t_{eq}\\
 \Psi_<(t,t_i) & = & {3\over 5}\Delta v \hspace{1mm}t_0^{2\over 3} t_i^{{1\over
2}}t^{2\over
3}t_{eq}^{-{5\over 6}}(1+\ln({t_{eq}\over t_i})),\hspace{5mm} t_i<t_{eq}
\end{eqnarray}

This result may be used to calculate several interesting quantities:
\begin{enumerate}
\item
 The large scale peculiar velocities predicted by the model.
\item
The nonlinear mass of objects seeded at an initial  time $t_i$.
\item
The mass function (for this we will also need the scaling solution
for monopoles).
\end{enumerate}
Here we find the predicted peculiar velocities and use this
result to normalize the single free parameter of the model $\eta$.
\par
It is easy to show using (11) that the predicted peculiar velocity caused by a
single monopole,
coherent on a scale q is \begin{equation}
v_>(t_0)={\dot\Psi}_>(t_0)={2\over 5}(\Delta v)({t_0\over
t_i})^{1\over 3}={60\over{\nu^{1\over 3}}}(G\eta^2)_6 v_m \gamma_m
q_{50}^{-1}km/sec
 \end{equation}
where $q_{50}$ is the comoving scale in units of 50$h^{-1}$Mpc, $\nu\geq 1|$
is the number of monopoles per horizon volume in the scaling solution
($t_i^{\rm com}=\nu^{1\over 3} q$) and $(G\eta^2)_6$ is $G\eta^2$
in units of $10^{-6}$. It may be shown using arguments similar to those in
Ref.\cite{tv91} that the
effect of all later monopoles is to produce an rms velocity larger by a factor
of 1.64 than the
result (13). Typically $\nu^{1\over 3}\simeq 1$\cite{rb90}. Demanding
$v_>(t_0)\vert_{q_{50}}\simeq
600km/sec$ we find $(G\eta^2)_6 v_m \gamma_m\simeq 10$ implying, for $v_m
\gamma_m\simeq 1$ a
symmetry breaking scale $\eta$ for the monopole forming phase transition
$\eta\simeq 3\times
10^{16}GeV$. This value of $\eta$ is consistent with the gravitational wave
background constraints
obtained in Ref.\cite{k91}  $\eta\leq 5\times 10^{16}GeV$.
\section {\bf Nonlinear Structures}
\par
The solution (11) , (12) describes how the expansion of $r_{\perp}(t)$ is
slowed down gravitationally due to the monopole induced velocity perturbation.
The Zeldovich approximation used in deriving it can be used until the shell
described by $r_{\perp}(t)$ turns around i.e. $\dot r_{\perp}(q,t)=0$ The
condition $\dot
r_{\perp}(q_{nl},t)=0$ determines the thickness $q_{nl}$
of the cylindrical
overdensity formed by the initial velocity perturbation. Using (5) it is easy
to
show that $q_{nl}(t,t_i)=2\Psi(t,t_i)$. The nonlinear scale $q_{nl}$ however
can
not grow beyond the comoving scale of the initial velocity perturbations
$t_i^{\rm com}$. Therefore, the correct expression for the turnaround scale at
the
present time $t_0$ is $q_{nl}(t_0,t_i)={\rm min}(t_i^{\rm com},2\Psi(t_0,t_i))$
or,
defining $t_*^{\rm com}=2\Psi(t_0,t_*)$ we have:
 \begin{eqnarray}
q_{nl}(t_0,t_i) & = & t_i^{\rm com},\hspace{5mm} t_i<t_*\\
q_{nl}(t_0,t_i) & = & 2\Psi(t_0,t_i),\hspace{5mm} t_i>t_*
\end{eqnarray}

The form of $q_{nl}(t_0,t_i)$ is shown in Figure 1 for $t_*>t_{eq}$. Notice
that for $t_i\leq t_*$ it
is $t_i^{\rm com}$, not $2\Psi(t_0,t_i)$, that determines the scale $q_{nl}$.
 It is straightforward to
calculate $t_*$ and show that
 $t_*=t_{eq}f(\lambda)$ where
$\lambda=0.4 (G\eta^2)_6 h^2 v_m \gamma_m$ and

\begin{eqnarray*}
 f(\lambda) & = & {\rm e}^{{\lambda-1}\over \lambda},\hspace{5mm} \lambda<1\\
 f(\lambda) & = & \lambda^{3\over 2}, \hspace{5mm} \lambda>1
\end{eqnarray*}


In what follows we will use the normalization of the previous section and
assume
$\lambda=4 h^2>1$
\par The nonlinear mass corresponding to structure originating at time $t_i$
may be written as
\begin{equation}
M(t,t_i)=\pi q_{nl}^2(t,t_i) v_mt_i^{\rm com}{\bar \rho}(t_0)
\end{equation}

For $\lambda>1$ we have $t_*>t_{eq}$ and using (14), (15) and (16) we obtain
\begin{eqnarray}
M(t_0,t_i) & = & {v_m \over {6G}}({t_i\over t_{eq}})^{1\over 2}t_i,
\hspace{5mm} t_i<t_{eq}\\
M(t_0,t_i) & = & {v_m \over {6G}}t_i, \hspace{5mm} t_{eq}<t_i<t_{*}\\
M(t_0,t_i) & = & {6\over 25}{{(\Delta v)^2 v_m}\over G}({t_0\over t_i})^{1\over
3} t_0, \hspace{5mm}
t_i>t_{*}
\end{eqnarray}
For $t_i<t_*$, $q_{nl}(t_0,t_i)\simeq t_i^{\rm com}$ and therefore all matter
goes nonlinear on
structures seeded before $t_*$. However, M$(t_0,t_i)$
is an increasing function of $t_i$ and therefore structures seeded at $t_i<t_*$
can
form by accreting
structures seeded before $t_i$. On the other hand, for $t_i>t_*$, M$(t_0,t_i)$
decreases with $t_i$
and since all matter has gone nonlinear in structures seeded before $t_*$,
monopoles after $t_*$ can
not produce nonlinear structures by today but can only induce large scale
velocity f1ows.
\par
The important point to notice in Figure 1 is that the distinguished scale
on which the largest and most prominent structures form is $t_*$
i.e. a scale which is
larger than $t_{eq}^{\rm com}$ by a factor $\lambda^{1\over 2}$. Thus the
characteristic scale
of dominant structures in the model is
\begin{equation}
t_*^{\rm com}=\lambda^{1\over 2} t_{eq}^{\rm com}=35 h^{-1} Mpc
\end{equation}

while the corresponding mass is
\begin{equation}
 M(t_0,t_*)\simeq 2\times 10^{16}h^{-1}M_\odot
\end{equation}
 where we have used the normalization obtained in the previous section
$\lambda=4h^2$.

\par
It is straightforward to now obtain the mass function predicted
in the model using the additional input of the monopole scaling solution.
Since the number of monopoles per Hubble volume per conformal horizon $\tau_i$
is ${{dn}\over {d\tau_i}}={\nu\over {\tau_i^4}}$ with $\nu\simeq O(1)$, it is
easy to
see that the number density of nonlinear structures with mass larger than M is
given by
 \begin{equation}
n_{>M}=\int_M^{M_*}dM{{\nu{d\tau_i\over dM}}\over {\tau_i^4(M)}}
\end{equation}
where M$(t_0,t_i)$ was obtained in (17)-(19).
 It is straightforward to compute the integral (22) to find $n_{>M}$. The
result is
shown in Figure 2 and may be written as  \begin{equation} n_{>M}^{\rm
model}={{\nu v_m}\over {144 G
t_0^2}}(M^{-1}-M_*^{-1})\Theta(M_*-M)  \end{equation}
We also show the Schechter mass function (obtained from the Shechter luminocity
function\cite{es83}
with a mass to light ratio of $100 {M_\odot \over L_\odot}$) in Figure 2 for
comparison with the
monopole result.   Clearly the model predicts a cutoff at the mass scale $M_*$.
This cutoff however
is much larger than the galactic mass and should be distinguished from the
observed cutoff in the
Schechter function. The Schechter function cuttof is expected to naturally
emerge when effects of
hydrodynamics are taken into account as was the case in Ref. \cite{t89}. In
particular\cite{ro77},
cooling effects may provide the cutoff since the gas on larger scales can not
cool in one Hubble
time.
 As shown in Figure 2 the magnitude of the predicted mass function in the
galactic mass range
is in
reasonably good agreement with the Schechter function given the assumptions
involved in the
calculation. The slope however is $-1$ for the monopole model but only $0.3\pm
0.1$ for the observed
Schechter function. This potential problem of the model could be resolved by
considering hot dark
matter or by introducing bias in the model.

\par
 We have therefore presented a study of a model for large scale structure
formation based on
primordial perturbations created by global monopoles and nonrelativistic dark
matter
(cold dark matter). The model has several encouraging features and clearly
deserves
further study. The model's single free parameter was fixed by demanding
agreement with the observed
large scale velocity f1ows. The obtained value was consistent with
gravitational wave background
bounds. The characteristic scale of dominant structures in the model is larger
than the comoving horizon at $t_{eq}$ implying that the model has significant
 power on large scales.
 The main distinct feature of the monopole wakes
considered here when compared to the cosmic string wakes is their generic
cylindrical geometry to be
compared with the planar geometry of string wakes.  In addition, the global
nature of the field in
the monopole model implies the existence of larger correlation length in the
scaling solution
compared to the case of cosmic strings. Thus the parameters of the scaling
solution in the case of
monopoles favor the formation larger structures than cosmic strings. \par
One of the potential problems of the model is its prediction that all mass has
gone
nonlinear at the present time $t_0$. This problem, which also appears in the
case of cosmic strings with cold dark matter  could be resolved by considering
relativistic particles as the dark matter (hot dark matter). Work in this
direction is in progress. The consideration of hot dark matter could also
soften
the slope of the mass function which is predicted to be steeper than the
Schechter function for galaxies.
\par
The global monopole model for structure formation predicts a distinct signature
on the microwave
backgroud. In particular, a global monopole moving perpendicular to the line of
sight of the
observer will produce a temperature dipole in the microwave sky i.e. a hot-cold
spot pair. The
detailed form, magnitude and distribution of this signature will be presented
in a separate
publication\cite{lp92}.  \\

{\bf Acknowledgments:}\\
I would like to thank Robert Brandenberger and Tanmay Vachaspati for
interesting
conversations. This work was supported by a CfA Postdoctoral Fellowship.\\
\\
\centerline{\Large \bf Figure Captions}
{\bf Figure 1}: The nonlinear mass $q_{nl}$ vs time of initial perturbation
$t_i$.\\
{\bf Figure 2}: The predicted mass function superposed with the observed
Schechter function
(obtained by using a fixed mass to light ratio).
 
\end{document}